# IMAGING METABOLICO AL CARBONIO 13 IPERPOLARIZZATO: STATO DELL'ARTE

Stefania Nicolosi[a]

[a] Laboratorio Tecnologie Oncologiche (LATO), C.da Pietrapollastra, Cefalù, Italy

## 1 Introduzione

La Risonanza Magnetica per Immagini (MRI) può essere considerata una delle più efficaci tecniche di diagnostica clinica sia in campo medico che biomedico, poichè permette di ottenere immagini anatomiche del corpo nelle sue varie parti e informazioni funzionali, come il flusso, la perfusione, la diffusione, il moto, ecc [1]. Essa si basa sull'interazione dei nuclei atomici con numero quantico magnetico di spin diverso da zero con il campo magnetico esterno e, in teoria, potrebbe essere sfruttata per l'analisi di diversi nuclei, come $^1H$, $^{13}C$, $^{31}P$ (endogeni) o $^{129}Xe$, $^3He$ (esogeni). In pratica, invece, l'uso clinico dell'MRI è ristretto all'analisi di $^1H$ che, rispetto agli altri, è presente in concentrazioni maggiori nell'organismo umano. L'intensità del segnale NMR dipende infatti dalla concentrazione dei nuclei attivi e dalla loro polarizzazione, definita, per nuclei con I = 1/2, come:

$$P = \frac{N^+ - N^-}{N^+ + N^-} \quad (1)$$

dove $N^+$ e $N^-$ rappresentano il numero degli spin con le due diverse orientazioni. All'equilibrio termodinamico la precedente equazione si riduce a

$$P = \tanh\left(\frac{\gamma \hbar B_0}{2 k_B T}\right) \quad (2)$$

dove $\gamma$ è il rapporto giromagnetico, $\hbar$ la costante di Plank, $B_0$ il campo magnetico esterno, $k_B$ la costante di Boltzmann e T la temperatura assoluta. All'equilibrio termodinamico la differenza di popolazione dei livelli e di conseguenza la polarizzazione P è molto bassa e ciò comporta una bassa intensità del segnale NMR. Ad esempio, per un campo magnetico di 1,5 T la polarizzazione dei protoni dell'$^1H$ è $5*10^{-6}$, mentre quella del $^{13}C$ è $1*10^{-6}$. La qualità diagnostica delle immagini MRI dipende dalle diverse intensità di segnale emessi da tessuti adiacenti. I vari tessuti vengono differenziati nell'immagine sulla base dei tempi di rilassamento ($T_1$ o $T_2$) che li caratterizzano o della densità protonica [1-2].

La differenza relativa dell'intensità ($I_A$ e $I_B$) dei segnali di due regioni adiacenti nell'immagine viene chiamata contrasto ed è data da:

$$C = \frac{I_A - I_B}{I_A + I_B} \quad (3)$$

Maggiore è la differenza fra le intensità, migliore sarà il contrasto C. Esso è determinato da numerosi fattori, alcuni intrinseci al campione (densità protonica, $T_1$ e $T_2$ dei protoni dell'acqua tissutale), altri di natura strumentale (sequenza usata, somministrazione di agenti di contrasto). L'intensità del segnale è proporzionale alla combinazione di due fattori: la differenza numerica tra le popolazioni dei nuclei nei due stati ($N^+ - N^-$) e l'energia Zeeman di transizione fra gli stessi, che come noto è proporzionale al campo statico applicato e al rapporto giromagnetico. In ambito medico un elevato contrasto è un requisito essenziale per ottenere immagini di qualità. Per migliorare la qualità dell'immagine, sia in termini di un maggiore rapporto segnale/rumore sia di migliore contrasto, si utilizzano agenti di contrasto paramagnetici costituiti da ioni $Gd^{3+}$, $Fe^{3+}$, $Mn^{2+}$ che inducono una variazione locale del campo magnetico e, quindi, dei tempi di rilassamento senza che vi sia una modificazione della polarizzazione [2]. L'uso di mezzi di contrasto in genere allunga i tempi di esecuzione di un esame di risonanza magnetica. Questi, invece, verrebbero notevolmente ridotti aumentando direttamente l'intensità del segnale del substrato. Ciò può essere ottenuto tramite un aumento di polarizzazione del substrato, ossia creando uno stato di non equilibrio dove la differenza di popolazione tra i livelli viene aumentata (Fig. 1). Questo stato si dice iperpolarizzato e negli ultimi anni sono state studiate e messe a punto diverse tecniche per raggiungere tale scopo.

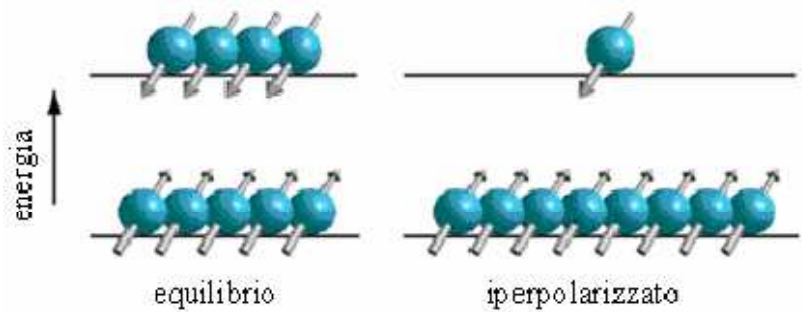

*Fig. 1*: Differenza di popolazione dei livelli energetici indotti dal campo magnetico nel caso di equilibrio termico e nello stato iperpolarizzato.

Gli agenti di contrasto iperpolarizzati agiscono quindi in maniera molto differente da quelli classici paramagnetici, in quanto, invece di modificare il segnale dovuto ai protoni circostanti, diventano essi stessi sorgente di segnale [3-4].
Per poter esser funzionali tali agenti di contrasto devono essere caratterizzati da un elevato tempo di rilassamento longitudinale $T_1$, perchè lo stato iperpolarizzato deve ovviamente conservarsi il più a lungo possibile dopo l'iniezione nel sistema circolatorio del paziente.

**2 Tecniche di iperpolarizzazione**

Sono noti diversi metodi per ottenere molecole iperpolarizzate:

*Brute Force* [3-4]
Dall'equazione (2) discende immediatamente che un incremento di polarizzazione può essere ottenuto aumentando il campo magnetico e diminuendo la temperatura.
Raffreddando, ad esempio, il campione alla temperatura dell'elio liquido, 4 K, in un campo magnetico di 20 T, la polarizzazione viene aumentata di un fattore mille. In realtà, queste condizioni sperimentali estreme sono difficili da raggiungere e il processo è comunque molto lento e richiederebbe l'uso di altri agenti di rilassamento che dovrebbero essere separati alla fine del processo dalle molecole iperpolarizzate. Questi fattori e i costi elevati limitano molto l'applicabilità del metodo.

*Optical Pumping* [5-6]
Questo metodo è applicabile solo ai gas $^{129}$Xe e $^3$He. Il gas nobile viene miscelato con i vapori di un metallo alcalino (*spin-exchange optical pumping*), o con i vapori di atomi metastabili (*metastability optical exchange*), e irraggiato con luce laser di opportuna frequenza polarizzata circolarmente. I metalli alcalini o gli atomi metastabili assorbono il momento angolare della luce laser e vengono polarizzati. Successivamente, la polarizzazione viene trasferita agli atomi del gas nobile tramite il processo di spin exchange, cioè in seguito agli urti atomici.
Solitamente nello *spin-exchange optical pumping* viene usato il Rb: i suoi vapori vengono condensati alla fine del processo e il gas nobile può essere estratto criogenicamente. In un campo magnetico, la transizione elettronica $S_{1/2} \rightarrow P_{1/2}$ degli atomi di Rb può essere indotta da luce laser polarizzata circolarmente e caratterizzata da una lunghezza d'onda di 795 nm, in grado di generare una transizione del Rb dallo stato fondamentale allo stato +1/2 o -1/2. La polarizzazione elettronica del Rb può poi essere trasferita agli atomi di gas nobile sia attraverso la formazione di legami deboli di Van der Waals sia tramite collisioni atomiche, secondo un processo multistep: il vapore di Rubidio, ad una pressione circa 1 ppm rispetto a quella dello Xe, viene polarizzato e le collisioni tra gli atomi di $^{129}$Xe e quelli di Rb* (Rb polarizzato) trasferiscono la polarizzazione al gas nobile:

$$Rb + h\nu \rightarrow Rb^*$$

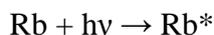

$$Rb^* + {}^{129}Xe \rightarrow Rb + {}^{129}Xe^*$$

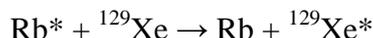

Il gas viene così polarizzato al 25% in 5-20 minuti a seconda delle condizioni ed è in grado di generare segnali NMR intensificati fino a 105 volte. Nel *metastablity optical pumping* invece si usa un plasma di atomi di $^3$He: la polarizzazione viene trasferita più velocemente, ma la tecnica è limitata solo alla produzione di $^3$He. In un sistema a bassa pressione di $^3$He, circa 1-2 mbar, si generano atomi metastabili con configurazione elettronica $^3S_1$ tramite una debole scarica elettrica. Usando luce laser polarizzata circolarmente a 1083 nm viene indotta la transizione $^3S_1 \rightarrow {}^3P_0$. Grazie al forte accoppiamento iperfine, i nuclei degli atomi metastabili si polarizzano e quando tali atomi collidono con un altro atomo non polarizzato nello stato elettronico fondamentale, si verifica un'alta probabilità di scambio della metastabilità.

L'atomo metastabile e quello fondamentale si scambiano la loro configurazione elettronica, mentre la polarizzazione nucleare rimane inalterata. La collisione genera così un atomo polarizzato allo stato fondamentale e uno non polarizzato in uno stato metastabile, che può andare incontro al processo di pompaggio ottico.

In entrambe le tecniche i gas iperpolarizzati sono compressi in celle di vetro prive di ferro, in modo da prevenire il rilassamento e mantenere la polarizzazione il più a lungo possibile.

*DNP: dinamic nuclear polarization* [3-5, 7]
Tale metodo sfrutta la forte polarizzazione degli elettroni, garantita dall'alto valore di γ che li caratterizza, rispetto alla polarizzazione nucleare in presenza di un campo magnetico relativamente basso. In condizioni moderate, 1 K e 3 T, la polarizzazione dei nuclei $^{13}$C è minore dello 0.1%, mentre quella degli elettroni è maggiore del 90%. Tale polarizzazione elettronica può essere trasferita nei solidi ai nuclei tramite l'applicazione di una radiofrequenza, quella di risonanza dell'elettrone, promuovendo transizioni flip-flop capaci di allineare gli spin nucleari tramite steps sequenziali. Il più lento processo di rilassamento nucleare rispetto a quello elettronico assicura il mantenimento dell'allineamento degli spin durante il processo e grazie al $T_1$ dell'elettrone estremamente corto, ~1 μs, esso rapidamente rilassa e viene nuovamente iperpolarizzato. Il processo si verifica anche a concentrazioni molto basse di elettroni disaccoppiati presenti nel campione.

In pratica, il materiale che deve essere polarizzato viene dopato con una specie radicalica stabile, di solito un nitrossido o un radicale tipo triarilico, e posto in un opportuno campo magnetico a bassa temperatura. La soluzione viene poi congelata e irradiata con opportuna radiofrequenza. Dopo che il trasferimento di polarizzazione è avvenuto, l'irraggiamento con radiofrequenze viene interrotto, il campione solido è sollevato sopra il livello dell'elio liquido e rapidamente riscaldato, solitamente tramite dissoluzione in acqua calda, ancora all'interno del campo magnetico. Viene poi rapidamente trasferito per l'osservazione o l'iniezione nel paziente. È stato dimostrato che con il metodo della dissoluzione in acqua calda si conserva in buon livello di polarizzazione.

In letteratura sono riportati molti esempi di molecole polarizzate tramite DNP e alcuni di questi verranno trattati nei prossimi paragrafi e nel prossimo capitolo. Sebbene questo metodo sia un valido strumento per l'MRI e MRS, soprattutto per l'alta applicabilità a composti solubili in acqua, non tossici e con interesse biologico, richiede tempi per raggiungere il massimo della polarizzazione molto lunghi (fino a 9 ore) e ciò ne limita le applicazioni.

*PHIP: parahydrogen induced polarization* [3-5, 8-9]

L'accoppiamento scalare e l'effetto Overhauser possono far avvenire il trasferimento della polarizzazione dall'idrogeno agli etero nuclei vicini nel prodotto di para-idrogenazione. L'iperpolarizzazione eteronucleare è maggiore se dovuta all'accoppiamento scalare e il fenomeno è stato trattato molto in letteratura, sia dal punto di vista teorico sia analizzando diverse molecole organiche insature. Fra queste citiamo a titolo di esempio la polarizzazione del $^{13}$C carbonilico dell'acetilendicarbossilato para-idrogenato, prima molecola para-idrogenata ad essere stata usata come agente di contrasto in MRI. Altri esempi verranno trattati nei prossimi paragrafi e nel prossimo capitolo.

Il metodo PHIP presenta come vantaggio principale rispetto al DNP il fatto di essere molto veloce, permettendo produzioni successive ripetute di 1-10 ml di soluzione $^{13}$C iperpolarizzata da iniettare ogni 2 minuti. Presenta però più limitazioni per quanto riguarda la scelta dei substrati da iperpolarizzare, dato che devono essere in grado di dare una reazione di idrogenazione veloce, e per l'eliminazione del catalizzatore e del solvente, in genere tossici per l'organismo.

## 3 Imaging con gas nobili iperpolarizzati

I primi esperimenti *in vivo* con gas nobili iperpolarizzati risalgono a metà degli anni novanta, precisamente nel 1994, quando ricercatori delle università di Stony Brook e Princeton (USA) sperimentarono l'uso di $^{129}$Xe iperpolarizzato per ottenere immagini MRI per gli studi sulla respirazione [5-6]. Lo Xeno infatti, oltre ad essere un anestetico generale molto sicuro, è in grado, se polarizzato, di fornire immagini angiografiche e dei polmoni. Fare un'immagine con un gas iperpolarizzato è molto vantaggioso perché viene notevolmente ridotto il rumore di fondo e si ottengono immagini molto più chiare [3-6]. Sia $^{129}$Xe che $^{3}$He, ad esempio, sono stati usati per studiare i polmoni, un'area particolarmente difficile da studiare con la tradizionale tecnica MRI. Infatti, anche alla fine dell'espirazione, la densità totale di nuclei $^{1}$H è solo 0.30 g/cm$^{3}$, una concentrazione troppo bassa per poter registrare immagini con una buona risoluzione. Invece, introducendo $^{129}$Xe polarizzato nei polmoni di un topo attraverso la trachea è stata ottenuta velocemente un'immagine dei polmoni e del cuore dell'animale con una risoluzione maggiore [10]. Lo $^{129}$Xe viene trasferito rapidamente dai polmoni al sangue e da qui ai tessuti, concentrandosi nelle componenti proteiche e lipidiche. Dato che il $T_1$ è abbastanza elevato, variando nei tessuti dai 15 ai 40 secondi, ed è dello stesso ordine di grandezza del tempo di trasferimento del gas polarizzato dai polmoni ai tessuti, è possibile ottenere immagini non solo dei polmoni ma anche del sistema circolatorio, del cervello o di altri organi vitali dell'organismo (Fig. 2.2).

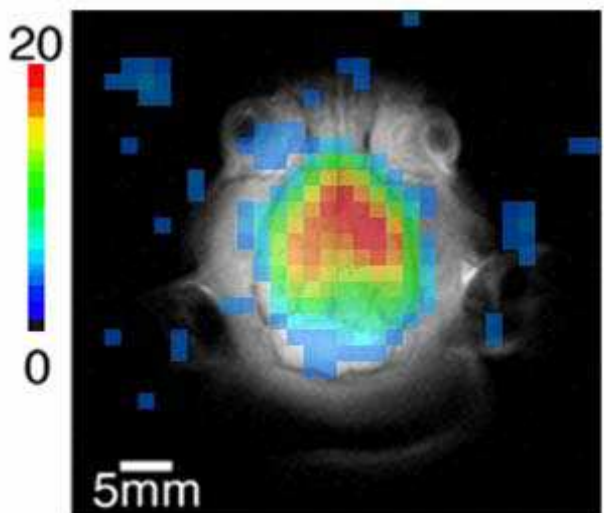

*Fig. 2.2*: In grigio è riportata l'immagine Spin Echo del cervello di un topo, a colori è sovrimposta l'immagine 32x32 pixels del cervello ottenuta dopo la respirazione di $^{129}$Xe iperpolarizzato per circa 40 secondi. Il colore rosso indica una concentrazione maggiore di $^{129}$Xe, mentre il blu una minore. Come si può notare il segnale dello 129Xe proviene principalmente dall'interno del cervello, mentre nel cervelletto la sua concentrazione è molto bassa.

## 4 Agenti di contrasto iperpolarizzati

L'uso di gas nobili iperpolarizzati è comunque limitato allo studio dei polmoni e ad un numero ristretto di studi di perfusione. Il recente sviluppo invece di metodi di iperpolarizzazione del nucleo $^{13}$C ha aperto un nuovo campo di applicazioni *in vivo* [3,4].

L'iniezione di una soluzione iperpolarizzata potrebbe essere utilizzata per visualizzare parti del sistema vascolare, mappare parametri fisiologici tramite perfusione e definire sonde per vie metaboliche. Infatti, gli agenti di contrasto $^{13}$C possono essere distribuiti in tutti i tessuti dell'organismo, garantendo molte più applicazioni pratiche. La possibilità di iperpolarizzare molecole marcate in $^{13}$C permette di superare il problema della sua bassa abbondanza isotopica (1.1%) e quindi della bassa sensibilità. Inoltre, l'osservazione del nucleo $^{13}$C supera il problema del rilassamento estremamente veloce dell'$^{1}$H: le interazioni dipolari, infatti, fanno diminuire velocemente la polarizzazione del protone subito dopo l'iniezione della molecola iperpolarizzata, mentre lo stesso processo è più lento per il carbonio, soprattutto per piccole molecole, grazie al minore rapporto giromagnetico. Le molecole marcate in $^{13}$C hanno infatti tempi di rilassamento di svariati secondi, da 10 a più di 60. Molti di questi substrati, inoltre, sono facilmente usabili *in vivo*, grazie alla loro atossicità. La concentrazione di $^{13}$C nei tessuti è molto bassa, così il rumore di fondo è completamente trascurabile e non influenza l'immagine. Risulta interessante confrontare il rapporto segnale rumore per gli agenti di contrasto $^{13}$C con i convenzionali per $^{1}$H MRI, in modo da poter stabilire i potenziali successi di tale tecnica (Tab. 2.1) [11].

|  | $^{1}$H | $^{129}$Xe | $^{3}$He | $^{13}$C |
|---|---|---|---|---|
| **γ (MHz/T)** | 42.6 | 11.7 | 32.5 | 10.7 |
| **Polarizzazione** | $5*10^{-6}$ | 0.2 | 0.3 | 0.15 |
| **Concentrazione** | 80 | 0.01 | 0.002 | 0.2 |
| **Fattore di diluizione** | 1 | 60 | 12 | 12 |
| **SNR relativo[1]** | 100 | 2.3 | 9.5 | 127 |

*Tab. 2.1*: Confronto di vari parametri di diversi nuclei.

---

[1] Il rapporto segnale rumore è normalizzato rispetto ad una resa di 100 per l'$^{1}$H

Come si nota dalla tabella, il rapporto segnale/rumore per $^{129}$Xe disciolto in biotrasportatori e per $^{3}$He incapsulato in microbolle è un ordine di grandezza minore rispetto a quello dei classici agenti di contrasto $^{1}$H. Al contrario, invece, il più alto accumulo di $^{13}$C nel sangue garantisce un SNR comparabile con quello degli agenti di contrasto $^{1}$H per una polarizzazione del 15%. Tenendo conto che la polarizzazione potrebbe essere aumentata fino ad un valore di circa il 30% tramite l'ottimizzazione della tecnica di iperpolarizzazione, è chiaro che il SNR può aumentare di un ordine di grandezza. Il basso rumore di fondo, come già evidenziato, dipende dalla bassa abbondanza isotopica del $^{13}$C. L'intensità del segnale ottenuto è proporzionale alla concentrazione dell'agente di contrasto iniettato e per questo motivo la tecnica presenta similitudini con la PET (tomografia ad emissione di positroni) e la SPECT (tomografia computerizzata ad emissione di singoli fotoni), in cui viene registrata la radioattività del tracciante. La mancanza di un segnale di fondo è vantaggiosa per molte tecniche, come ad esempio le angiografie, dove si ricerca il più alto grado di contrasto tra i vasi e l'intorno. D'altra parte, però, senza un segnale di fondo l'interpretazione anatomica di tali immagini diventa problematica e in questi casi vengono poi sempre registrate immagini $^{1}$H classiche di riferimento, con uguale orientazione, in modo tale da poter successivamente sovrapporre le due diverse immagini ottenute.

## 4.1 Iperpolarizzazione del $^{13}$C indotta dal para-idrogeno

Una buona via per ottenere un'immagine MRI consiste nello sfruttare la polarizzazione che deriva dal para-idrogeno [8]. In questo caso, un substrato insaturo viene idrogenato con una miscela di idrogeno arricchita nella forma para e, dopo la rimozione del catalizzatore, il prodotto iperpolarizzato è iniettato nel paziente come agente di contrasto. Esso fluisce nel flusso sanguigno e sfruttandone i segnali NMR intensificati è possibile ottenere immagini angiografiche. Questa tecnica è vantaggiosa perché in linea di principio:
- è possibile variare chimicamente il substrato da idrogenare, in modo da aumentarne la biocompatibilità,
- il trasferimento di polarizzazione ad eteronuclei risolve il problema del rumore di fondo.

Ma ci sono molte difficoltà da superare:
- la ricerca di un catalizzatore efficiente per la reazione di idrogenazione, che favorisca il trasferimento di polarizzazione ad eteronuclei,
- la ricerca di un substrato che dopo la reazione generi una specie iniettabile nell'organismo (biocompatibile e solubile in mezzo acquoso),
- la rimozione del catalizzatore, che, essendo un complesso di metalli di transizione come Rh, Ir, Os, ecc, è molto tossico per l'organismo,
- l'eliminazione del solvente di reazione se non è biocompatibile. In genere si usa l'acetone che ha un LD50 orale di 5800 mg/kg (ratto).

Inoltre, prima di iniettare la molecola para-idrogenata nell'organismo, per poter registrare l'immagine, è necessario convertire l'alto ordine di spin di non equilibrio derivante dal para-idrogeno in un ordine di magnetizzazione longitudinale, in modo da avere segnali non più in antifase (che avendo una componente in assorbimento e una in emissione hanno intensità totale nulla), ma completamente in fase, o in emissione o in assorbimento. Questo può essere fatto in due modi diversi: o applicando un ciclo di campo al campione para idrogenato o irradiandolo con opportuna radiofrequenza. Il ciclo di campo consiste nel ridurre velocemente la forza del campo magnetico, passando dal campo magnetico terrestre, a cui avviene la reazione, ad un valore minore di 0.1 μT e successivamente riportarlo adiabaticamente, cioè lentamente, a campo magnetico terrestre. In condizioni di basso campo il sistema di spin protone-carbonio è portato ad un regime di forte accoppiamento, esaltando l'accoppiamento scalare, e, quando il campo viene aumentato lentamente, per steps successivi, le popolazioni di spin sono mantenute. Il risultato è uno spettro

NMR polarizzato $^{13}$C dove le transizioni permesse sono predominatamente in fase, corrispondenti ad una polarizzazione netta. L'ottimizzazione della procedura è stata raggiunta attraverso una simulazione dell'evoluzione del sistema di spin ed è differente per ogni molecola.

Questo metodo è stato applicato con successo la prima volta, usando un cilindro schermante di μ-metal, da Golman et al. sul prodotto di para-idrogenazione 40 dell'acetilendicarbossilato di metile [12] e dell'idrossietilpropionato di metile [13], ottenendo un'intensificazione del segnale $^{13}$C di $10^4$ volte rispetto alle molecole non polarizzate. Questi substrati sono stati poi usati come agenti di contrasto per angiografie in ratti. Secondo quanto riportato da Golman e Johannesson lo stesso risultato può essere conseguito usando un'opportuna sequenza di impulsi [14]. Questa consiste in una irradiazione continua sul campione durante la reazione di idrogenazione, in modo da minimizzare la perdite di ordine di spin. Terminata la reazione viene applicato un impulso 180x sul protone in modo da preparare il sistema ad una serie di impulsi 90x e 90y sul carbonio, efficaci per la conversione dell'ordine di spin del $^{13}$C in polarizzazione netta. Sono introdotti molti echi nella sequenza per correggere i difetti legati a disomogeneità di campo (Fig 2.3).

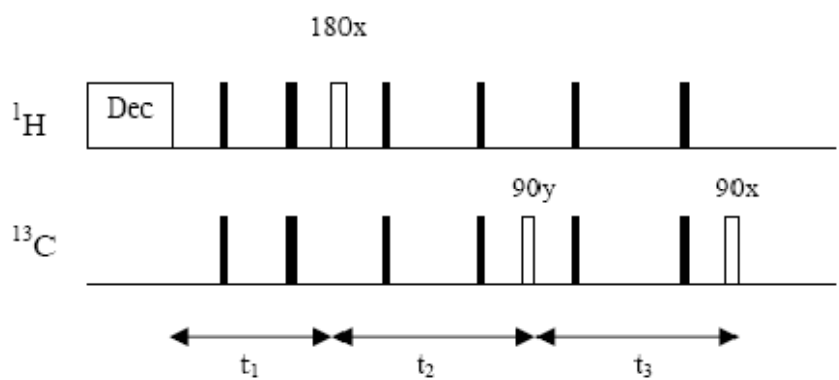

*Fig. 2.3*: Schema della sequenza di impulsi per portare in fase i segnali polarizzati. Dopo il disaccoppiamento, ci sono solo 3 impulsi "utili", in bianco, mentre gli altri, in nero, sono gli echi.

Per poter utilizzare la polarizzazione derivante dal para-idrogeno per acquisire un'immagine è necessario che il substrato, oltre ad idrogenarsi molto velocemente, abbia un lungo tempo di rilassamento longitudinale del nucleo $^{13}$C che dà origine al segnale iperpolarizzato. In genere, i valori riportati di $T_1$ e $T_2$ per molecole usate *in vivo* come agenti di contrasto $^{13}$C sono tipicamente 10-100 volte più lunghi di quelli dei protoni e in genere cadono tra i 20-45 secondi per il $T_1$ e i 2-5 secondi per il $T_2$ [3-5]. Il $T_1$ di una molecola organica dipende da svariati fattori [15], ma il meccanismo predominante è l'interazione dipolare tra il nucleo in esame e i protoni vicini (accoppiamento spin-spin). Per un atomo di carbonio che interagisce con *n* protoni la velocità di rilassamento dipolare R, cioè l'inverso del tempo di rilassamento longitudinale $T_1$, si esprime come:

$$R = \frac{1}{T_1} = n \left(\frac{\mu_0}{4\pi}\right)^2 \left(\frac{\hbar^2 \gamma_H^2 \gamma_C^2}{r_{C-H}^6} f(\tau_C)\right)$$

dove $\mu_0$ è la suscettività magnetica nel vuoto, $\gamma_H$ e $\gamma_C$ i rapporti giromagnetici di idrogeno (42.58 Mhz/T) e carbonio (10.71 Mhz/T), $\tau_C$ il tempo di riorientazione molecolare e $r_{C-H}$ la distanza dei due nuclei. L'efficacia quindi del rilassamento longitudinale dipolare intramolecolare dipende da parametri come: la natura dei due spin attraverso il loro valore di γ, dalla loro distanza spaziale, dal tempo di correlazione molecolare. Quindi, per avere un rilassamento lento dei nuclei $^{13}$C è necessario che r sia elevata. Il tempo di reorientazione molecolare è il tempo durante il quale il sistema conserva memoria dello stato precedente, prima che intervenga una variazione casuale dello stato. Per le interazioni dipolari, il moto che causa le fluttuazioni del campo è di tipo rotazionale e dipende da: dimensioni della molecola (molecole grosse hanno lungo $\tau_C$), temperatura (aumentandola $\tau_C$ diminuisce), energia di attivazione del processo di rotazione (diminuisce $\tau_C$) e viscosità del solvente (se elevata ostacola il moto molecolare e perciò aumenta $\tau_C$).

Oltre al meccanismo dipolare esistono altri meccanismi che contribuiscono al rilassamento totale della molecola, tra i quali:
- le interazioni scalari dovute all'accoppiamento indiretto J scalare con una specie che scambia rapidamente o con un nucleo che rilassa rapidamente,
- l'anisotropia del tensore di schermo (chemical shielding anisotropy, c.s.a.) legata al fatto che lo schermo elettronico nucleare, che genera un campo locale a cui è soggetto il nucleo, non è isotropo ma dipende dall'orientazione della molecola rispetto al campo magnetico esterno,
- l'interazione dipolare intermolecolare con nuclei di altre molecole,
- l'interazione dipolare paramagnetica con spin elettronici spaiati di altre molecole.

Quest'ultimo è un fattore importante. L'ambiente sanguigno è ricco di ossigeno che, essendo una molecola biradicalica, contribuisce in modo significativo al rilassamento del prodotto iperpolarizzato. Il $T_1$ del substrato idrogenato deve quindi essere molto elevato perché, dopo l'iniezione nell'ambiente biologico, l'interazione con l'$O_2$ ne causa un veloce rilassamento e quindi una veloce perdita della polarizzazione. Le prove di paraidrogenazione in tubo vengono infatti sempre fatte dopo la deossigenazione delle soluzioni, anche perché i catalizzatori di idrogenazione che normalmente si usano formano complessi stabili in presenza di ossigeno e si disattivano.

## 4.2 Iperpolarizzazione del $^{13}C$ tramite DNP

L'iperpolarizzazione DNP può essere attribuita a due meccanismi diversi:
- solid effect
- thermal mixing

I due meccanismi possono operare simultaneamente o separatamente. Nei paragrafi che seguono è riportata una breve descrizione di entrambi i meccanismi finalizzata a dare semplici criteri sperimentali per distinguere quale fra essi determina la DNP. Per una trattazione teorica completa si rimanda a (ABRAGAM).

### 4.2.1 Solid Effect

Consideriamo una coppia di spin costituita da uno spin elettronico S=1/2 e da uno nucleare I=1/2 immersi in un forte campo magnetico a bassa temperatura. Lo stato del sistema isolato può essere descritto in termini dei numeri quantici $m_S$ (±1/2) e $m_I$ (±1/2) associati alle componenti $S_Z$ e $I_Z$ degli spin i cui corrispondenti livelli energetici sono rappresentati in figura ?

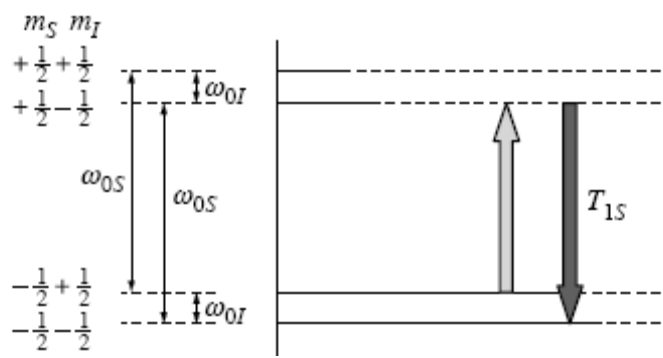

*Fig. :* A sinistra sono rappresentati i livelli energetici della coppia di spin….

Un impulso di frequenza ($\omega_{0S} - \omega_{0I}$) induce la transizione dallo stato ($m_S$=1/2, $m_I$=-1/2) allo stato ($m_S$=-1/2, $m_I$=1/2) e quindi determina un trasferimento di polarizzazione dall'elettrone al nucleo.
Questo processo è in grado di allineare un singolo spin nucleare. Dopo che l'equilibrio termico tra spin elettronico e reticolo si è ristabilito (tempo caratteristico $T_{1S}$) il processo può essere ripetuto per gli altri spins nucleari. Questo processo deve competere con l'interazione tra gli spins nucleari e il reticolo che tende a far disperdere la polarizzazione degli spins nucleari.
Un impulso di frequenza ($\omega_{0S} + \omega_{0I}$) ha l'effetto opposto, trasferisce, cioè, la polarizzazione elettronica positiva in polarizzazione nucleare negativa. L'effetto solido si osserva quando l'ampiezza della linea ESR è significativamente più piccata della frequenza $\omega_{0I}$. Sotto tali condizioni l'andamento della polarizzazione nucleare $P_I$ come funzione della frequenza $\omega_m$ dell'impulso dato è rappresentato nella figura seguente.

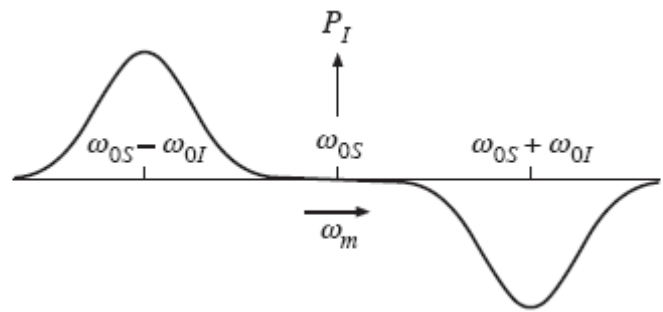

### 4.2.2 Thermal mixing

Il thermal mixing è un processo che induce la polarizzazione dinamica differente da quello appena descritto e può avvenire contestualmente o indipendentemente dal precedente. Contrariamente all'effetto solido in cui i nuclei sono polarizzati in un unico step per mezzo di una doppia transizione quantistica, la DNP indotta dal thermal mixing è un processo a due step di cui il primo è una singola transizione quantistica e richiede, in genere, meno potenza di quella richiesta per la realizzazione dell'effetto solido. Perché, inoltre, il thermal mixing sia efficiente occorre che l'ampiezza della linea ESR sia maggiore o eguale alla frequenza NMR. La concentrazione degli spins elettronici deve essere alta così che l'accoppiamento dipolare fra gli stessi sia sufficientemente elevato da poter ipotizzare che gli spins elettronici siano inizialmente caratterizzati da una loro temperatura (in genere differente da quella del bagno con cui interagiscono)(vedi abragam). Se queste condizioni sono soddisfatte il thermal mixing risulta il processo di polarizzazione dominante.
Il primo step, noto come *raffreddamento dinamico,* consiste nell'irradiare il sistema di interesse con radiazione elettromagnetica di frequenza $\omega_m$ diversa dalla frequenza di Larmor $\omega_{0S}$ degli spin elettronici così che, nel sistema di riferimento rotante, gli spins elettronici acquistino una temperatura $T_r$ diversa dalla temperatura $T$ del reticolo e a questa legata dalla relazione (vedi Redfield):

$$T_r = \frac{\Delta_0^2 + aD^2}{\omega_{0S}\Delta_0} T$$

dove $\Delta_0 = \omega_0 - \omega_m$, $D$ è il momento dipolare associato alla linea ESR e $a \approx 2$ è una costante numerica dipendente dal modello scelto. $T_r$ può essere svariati ordini di grandezza minore di $T$.
Il secondo step, noto come *thermal mixing in senso stretto*, rappresenta il raggiungimento dell'equilibrio termodinamico fra gli spins elettronici e gli spins nucleari. La temperatura di equilibrio raggiunta dipende dalle capacità termiche dei due sistemi di spin secondo. Risulta, quindi, evidente che maggiore è la concentrazione elettronica del sistema considerato più efficiente sarà il raffreddamento degli spin nucleari e una maggiore polarizzazione nucleare verrà indotta nel sistema.

Il processo descritto è schematizzao nella figura seguente

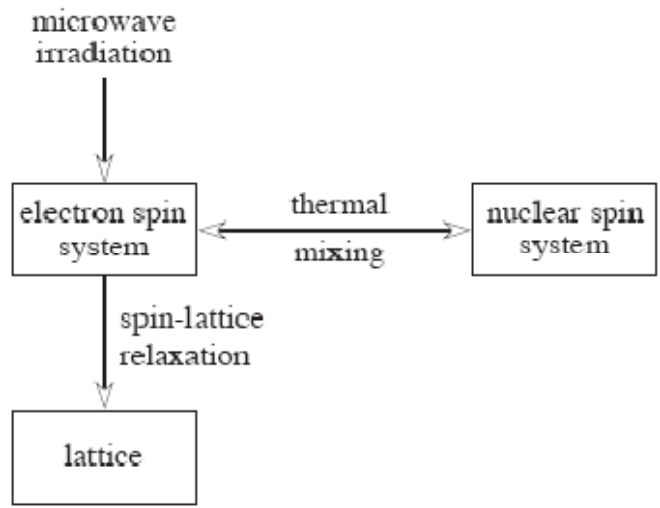

## 5 Sequenze di impulsi

Le sequenze classiche di impulsi che generalmente vengono applicate per ottenere un'immagine MRI richiedono molte scansioni e quindi, solitamente, sono necessari dai 4 ai 20 minuti. Per costruire una normale immagine $^1$H si deve applicare un'opportuna sequenza di impulsi e un certo gradiente di campo, che vengono ripetuti in successione appena il sistema rilassa. Il tempo quindi per ricavare un'immagine è troppo lungo se confrontato con il tempo di rilassamento della molecola iperpolarizzata, che rappresenta il principale fattore limitante. Dato che la polarizzazione decade esponenzialmente nel tempo, si perde velocemente la situazione di non equilibrio ed è necessario usare sequenze di impulsi in grado di registrare un'immagini in pochi secondi.
Ci sono essenzialmente due possibili vie per acquisire un'immagine di un mezzo iperpolarizzato:
- usando una serie di impulsi con un basso flip-angle (Fast Low Angle Shot, FLAS),
- usando una sequenza single shot (unica scansione).

Nel 1986 è stata introdotta da Jürgen Henning la sequenza RARE (Rapid Acquisition with Relazation Enhancement, anche detta Rapid Spin Echo RSE, Fast Spin Echo, FSE o Turbo Spin Echo, TSE) [16]. Essa permette di ottenere buone immagini in un tempo breve. Viene creato un treno di echi e ogni echo è codificato indipendentemente. Si basa su una sequenza Spin Echo multiple Echo (Fig. 2.4): il campione viene eccitato una sola volta, con impulso di 90° e dopo si applicano vari impulsi a 180°, che creano più echi che, a loro volta, generano una singola immagine. Il numero di echi che si possono generare dipende naturalmente dal tempo di rilassamento longitudinale della specie esaminata.

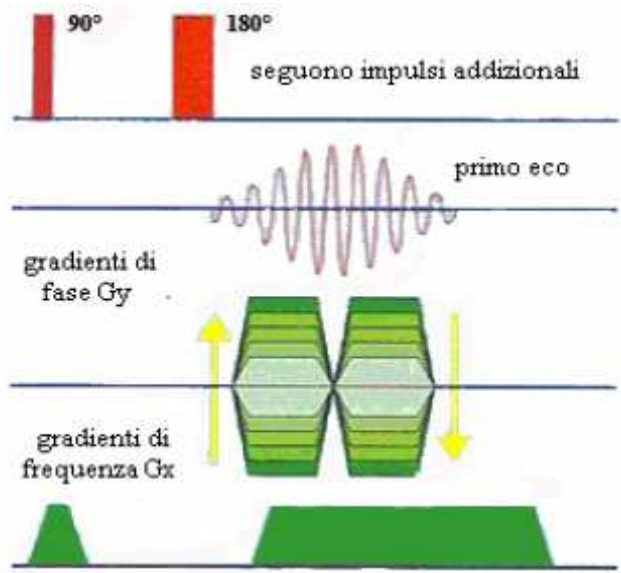

*Fig. 2.4*: Sequenza di impulsi RARE.

Anche la sequenza EPI (Echo Planar Imaging) sembra particolarmente utile dato che converte la magnetizzazione longitudinale iniziale in magnetizzazione trasversale con circa il 100% di efficienza, permettendo così di acquisire diverse immagini e di seguire la distribuzione dell'agente di contrasto. Ma la sequenza più usata è la trueFISP, Fast Imaging with Steady-state Precesion, con lunghi flip angle tra 160 e 180°, che permettono di riciclare la magnetizzazione trasversale da un ripetitore all'altro (Fig. 2.5) [11]. Questa è indicata in particolare per molecole in grado di rilassare lentamente, perché, in questo caso, possono essere usati ampi flip angles, in grado di bilanciare bene le disomogeneità di campo. Sebbene sia stata introdotta 20 anni fa, non è stata applicata per molto tempo a causa della mancanza di sistemi a forti gradienti, ma ora grazie all'alta intensità del segnale e alla velocità di acquisizione, questa sequenza è largamente sfruttata in molte applicazioni cliniche.

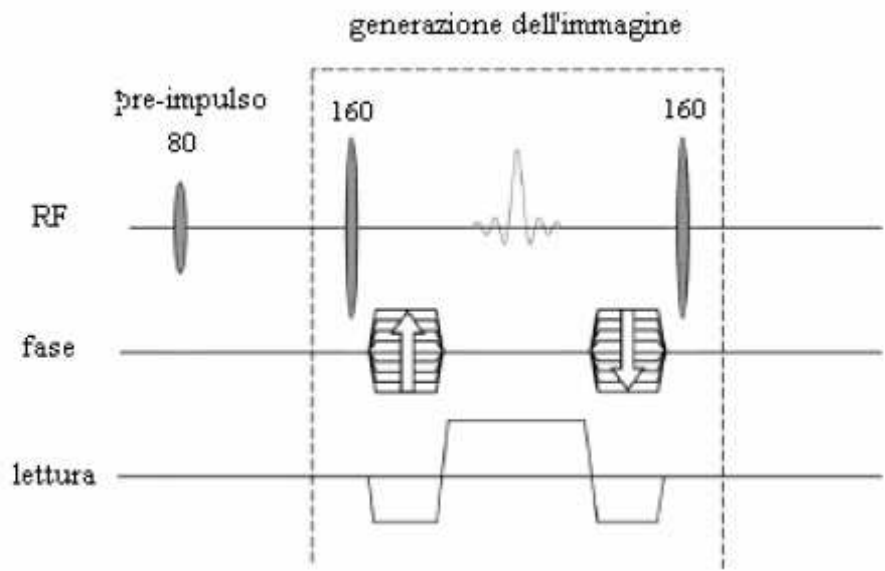

*Fig. 2.5*: Schema della sequenza trueFISP.

# 6. MRI 13C iperpolarizzato: applicazioni cliniche

Il comportamento farmacocinetico e farmacologico dell'agente di contrasto dipende ovviamente solo dalla molecola scelta, dato che l'iperpolarizzazione non cambia nessuna proprietà chimica o fisica della sostanza [3-5].

I primi esperimenti *in vivo* sono stati fatti usando sostanze non endogene e biologicamente inattive, in modo da produrre semplici angiografie. Per questi studi sono necessarie molecole con peso molecolare non troppo basso per evitare che la sostanza extravasi. È necessario però raggiungere un compromesso circa le dimensioni molecolari, dato che, per avere un T1 sufficientemente lungo, il peso molecolare dovrebbe essere solo dell'ordine di poche centinaia di unità di massa atomica. Successivamente si sono sperimentate sostanze endogene, come ad esempio l'urea, e recentemente la tecnica di imaging *in vivo* è stata centrata sullo studio di sostanze bioattive, in modo da poter registrare immagini in tempo reale dei metaboliti, come è stato dimostrato dallo studio dell'[1-$^{13}$C] piruvato, usato come marker metabolico da Golman et al. [4].

Nelle applicazioni *in vivo* i livelli di polarizzazione raggiunti sono del 20-30%, i valori di $T_1$ e $T_2$ sono rispettivamente di ~ 40 e 4 secondi, la concentrazione iniziale della soluzione iniettata tra 0.3 e 1.2 M e a causa della diluizione nel sistema vascolare si stima una concentrazione nell'organismo tra 2 e 40 mM, valore molto lontano dalla concentrazione protonica locale di circa 80 M.

Le potenziali applicazioni cliniche dei composti $^{13}$C iperpolarizzati sono:
- Immagini vascolari e angiografie
- Studi di perfusione
- Catheter tracking
- Imaging molecolare

## 6.1 Immagini vascolari e angiografie

In generale, l'angiografia è una rappresentazione dei vasi sanguigni o linfatici del corpo umano tramite tecniche invasive che prevedono l'introduzione per via cutanea di un catetere all'interno dei vasi e la generazione di immagini mediche dopo infusione di un opportuno mezzo di contrasto idrosolubile. Sono state sviluppate anche tecniche non invasive, come la tomografia computerizzata, l'ecografia e la risonanza magnetica. Questi esami sono fondamentali per individuare anomalie cardiopatiche, vasi ristretti o occlusi. La disponibilità di soluzioni $^{13}$C iperpolarizzate con tempi di rilassamento relativamente lunghi, rende possibile la registrazioni di immagini MRA (Resonance Magnetic Angiography) in tempo reale del sistema vascolare, con la possibilità di sfruttare tali agenti di contrasto per esaminare condizioni patologiche. Solo i vasi sanguigni contengono la sostanza iniettata, in grado di generare il segnale nell'immagine e, realizzando tutto in un tempo molto ridotto, si riduce la formazione di artefatti. Così facendo si possono ottenere immagini 2D di uguale, se non migliore, qualità di quelle 3D classiche $^1$H. Il primo esperimento riportato risale al 2001 da parte di Golman, Johannesson et al. [17]: essi registrarono un angiogramma completo di un ratto a 2.4 T usando una soluzione 150 mM in acetone del dimetilestere dell'acido maleico marcato in $^{13}$C, ottenuto dalla paraidrogenazione del dimetil estere dell'acido acetilendicarbossilico con una polarizzazione stimata *in vivo* di 0.3% e usando la sequenza di impulsi RARE, con un tempo complessivo di scansione di 0.9 s (Fig. 2.6). Il lungo tempo di rilassamento del carbonile marcato (75 secondi in acetone-d6 a 7.05 T), ha permesso di manipolare il prodotto e iniettarlo nella vena della coda del ratto con solo una piccola perdita di ordine di spin. L'immagine mostra chiaramente la vena cava e alcune ramificazioni, con una struttura fine visibile minore di 0.5 mm.

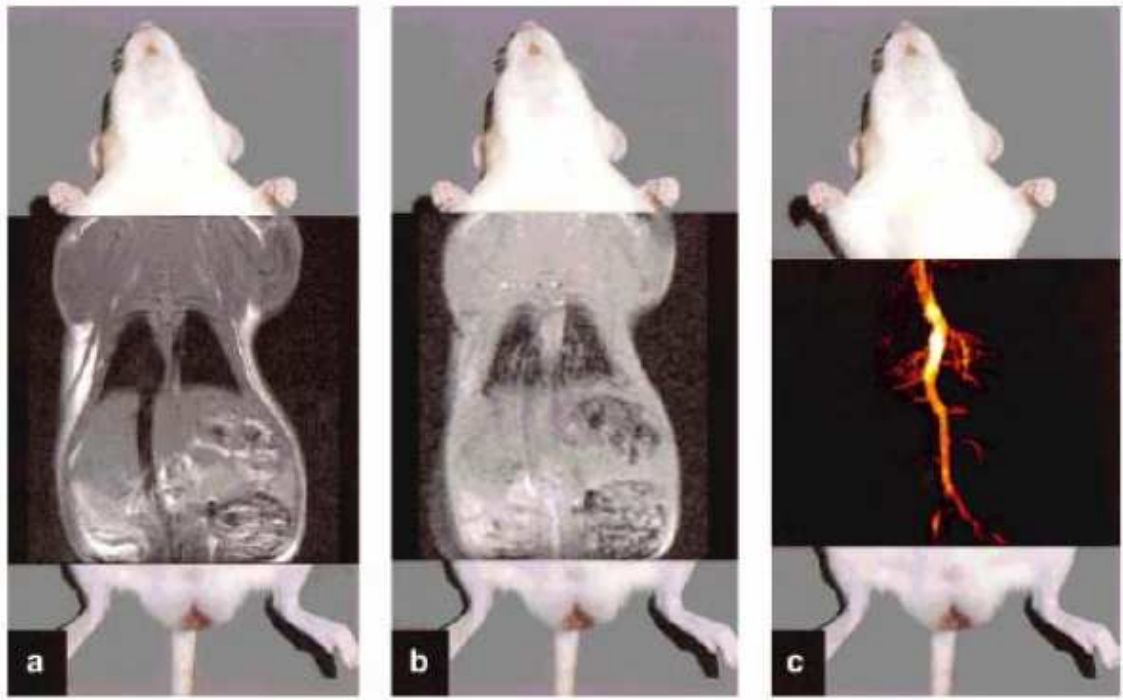

*Fig. 2.6*: Immagine di ratto registrata a 2.4 T: a) immagine 1H, b) immagine 1H con un tempo di scansione di 0.9 s, c) immagine $^{13}$C RARE in 0.9 s.

Il lavoro soffre comunque per la bassa polarizzazione (è stato usato para-idrogeno arricchito solo al 50%), e per il fatto che tutta la miscela di reazione, incluso il catalizzatore e l'acetone, sono stati iniettati nel ratto. Gli stessi autori hanno successivamente riportato un metodo per eliminare sia il catalizzatore che il solvente organico [3]: il catalizzatore ionico di Rh può essere rimosso tramite colonna a scambio cationico, mentre il solvente tramite distillazione spray flash dopo l'addizione di acqua alla miscela. Questi due processi sono comunque critici in quanto comportano una perdita parziale della polarizzazione del substrato. Anche il riempimento dinamico e l'estensione delle arterie coronariche possono essere visualizzati tramite l'iniezione di una sostanza 13C iperpolarizzata (Fig. 2.7) [4]. Un esperimento simile è stato eseguito su un maiale da Golman et al.: è stata selezionata una fetta sottile di 15 cm del cuore usando la sequenza bilanciata SSFP (steady state free procession) e si è registrata un'immagine ogni 422 ms, cioè ad ogni battito cardiaco. In alcuni casi l'alta velocità di flusso nel catetere origina degli artefatti di soppressione del segnale (frecce in Fig. 2.7 b).

Lo svantaggio principale nell'usare per angiografie il segnale $^{13}$C è il fatto che il basso rapporto giromagnetico rende difficile ottenere una grande risoluzione spaziale, a meno che i tempi di eco siano prolungati, ma ciò degrada la qualità dell'immagine.

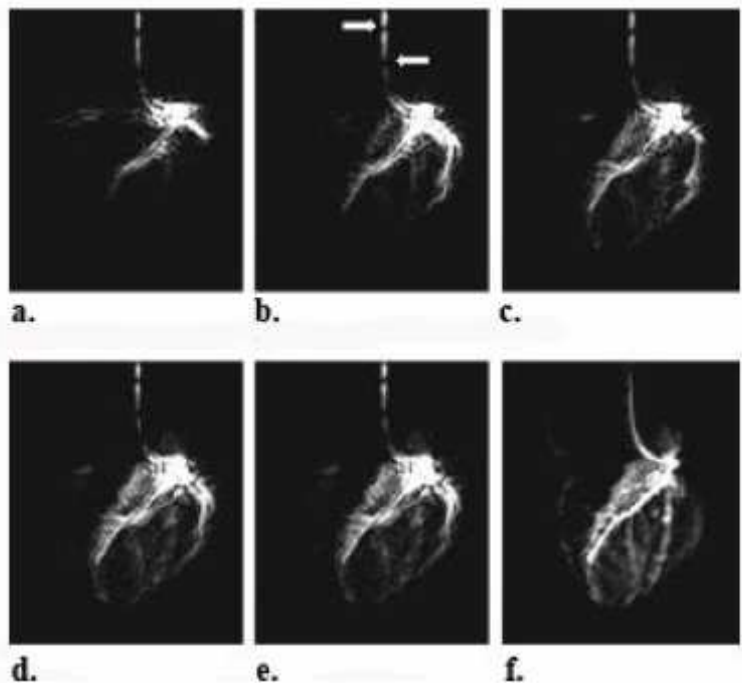

*Fig. 2.7*: Serie di angiogrammi ottenuti durante l'iniezione, attraverso un catetere posto nella arteria coronaria sinistra di un maiale, di una soluzione iperpolarizzata via DNP di (1-idrossimetil-1-13C-ciclopropil)-metanolo.

## 6.2 Perfusione

Nell'accezione più generale la perfusione indica il rilascio di sostanze nutrienti necessarie alle cellule per lo svolgimento delle loro funzioni vitali. In realtà, le tecniche convenzionali di misura della perfusione forniscono informazioni soltanto riguardo alla microcircolazione capillare, senza coinvolgere i meccanismi di compartimentazione delle sostanze. Da un punto di vista operativo, quindi, la perfusione viene definita come la quantità di sangue rilasciata da un voxel in un certo intervallo di tempo e viene misurata in $ml_{sangue}/minuti/g_{tessuto}$. Dato che essa avviene a livello capillare, ha proprietà fisiche notevolmente differenti da quelle del flusso coerente dei grandi vasi. Il moto perfusivo è infatti assimilabile ad un flusso microscopico incoerente, inteso come somma di molti flussi coerenti distribuiti in maniera casuale ed è detto moto pseudo-diffusivo (Fig. 2.8).

Alcuni parametri emodinamici cerebrali, come il flusso ematico celebrale (CBS), il volume di sangue celebrale (CBV) e il tempo di transizione medio (MTT), connessi alla perfusione, sono estremamente importanti per la diagnosi clinica di alcune patologie cerebrovascolari. Negli ultimi 25 anni diverse tecniche sono state messe a punto per la misura di queste quantità, come le metodiche nucleari PET e SPECT e alcune metodiche di MRI dinamico, che in genere offrono il vantaggio di non essere invasive e di possedere una più alta risoluzione spaziale e temporale. Esse sfruttano agenti di contrasto dinamici e le variazioni dei segnali ottenute in seguito all'iniezione degli agenti paramagnetici di Gadolinio sono analizzate sulla base della "bolus tracking theory"2.

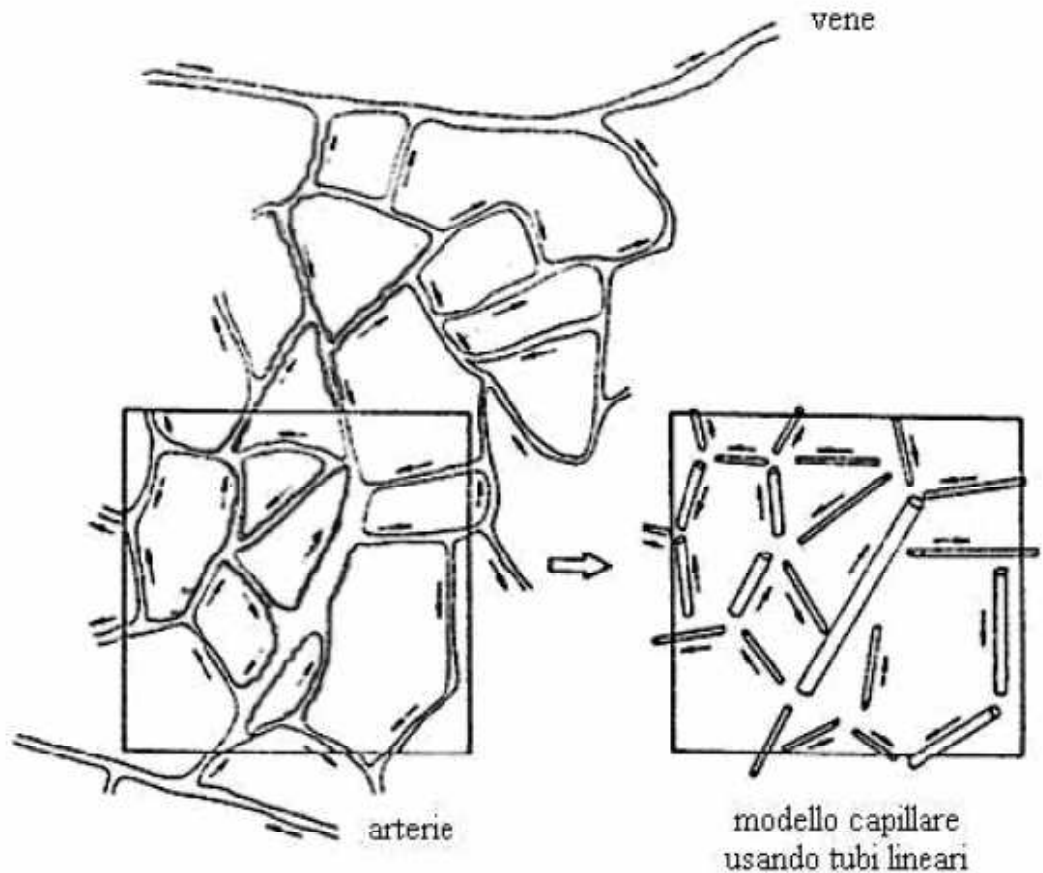

*Fig. 2.8*: Moto perfusivo.

In realtà, questa tecnica mostra diversi aspetti problematici qualora si vogliano avere informazioni quantitative assolute del CBS e del CBV, a causa della presenza di "artefatti vascolari". Infatti, si verificano diversi problemi relativi alla composizione vascolare dei tessuti (che influenza l'intensità del segnale), il diverso meccanismo di rilassamento dei protoni intra- ed extra-vascolari, la correlazione indiretta tra segnale e concentrazione dell'agente di contrasto e tutto ciò rende difficile la quantificazione della perfusione con i convenzionali marker. Allo stato attuale esistono diverse strategie NMR in grado di rilevare la perfusione cerebrale, tra cui:
- Dinamic susceptibility contrast (DSC): una sostanza paramagnetica viene iniettata come agente di contrasto esogeno,
- Arterial spin labeling (ASL): vengono marcati i protoni dell'acqua del sangue arterioso perturbandone la magnetizzazione con opportuni impulsi rf, trasformando così il sangue arterioso in un agente di contrasto endogeno, 2 È un metodo di inseguire gli spin mobili dopo averli targati, ad esempio dopo aver alterato localmente laloro magnetizzazione. Gli spin mobili sono registrati a tempi diversi dopo la marcatura in modo da mostrare dove il bolus (il piccolo volume di spin marcato) si è mosso nel piano dell'immagine.
- Contrasto BOLD (blood oxygen level dependent): basato sulla variazione della concentrazione locale di deossiemoglobina e dipendente dalla interazione tra i valori istantanei del flusso e del volume ematico celebrale e quelli di estrazione tissutale di ossigeno.

Negli ultimi anni è stato suggerito di usare l'aumento di segnale di molecole $^{13}C$ iperpolarizzate non solo per la visualizzazione del sistema cardiovascolare, ma anche per effettuare misure di perfusione [3-5]. In questo caso però la difficoltà maggiore nelle analisi quantitative è la perdita di polarizzazione durante il tempo dell'analisi, che produrrebbe un apparente più veloce decadimento del segnale. Per ovviare a ciò, possono comunque essere introdotte alcune modifiche nella teoria del bolus tracking, permettendo una determinazione abbastanza precisa dei parametri di perfusione. Le molecole studiate a tale scopo non attraversano la barriera sangue-tessuto cerebrale in maniera rilevante e quindi rimangono abbastanza confinate nel letto vascolare. Oltre al cervello è possibile studiare altri tessuti come cuore, rene, polmoni, dove i volumi di distribuzione dei traccianti sono più alti. Ad esempio, il bis-1,1-(idrossimetil)-1-$^{13}$C-ciclopropano-d$^8$, polarizzato tramite DNP, è stato usato per ottenere mappe di perfusione del miocardio di un maiale, iniettandolo sia tramite un catetere arteriale sia nella vena femorale, dove si è ottenuto un più basso SNR ma una migliore risoluzione spaziale. Studi sui polmoni di maiale sono stati condotti invece usando il 2-idrossietilacrilato marcato $^{13}$C e polarizzato al 20-30 % con metodo PHIP (Fig. 2.9) [3].

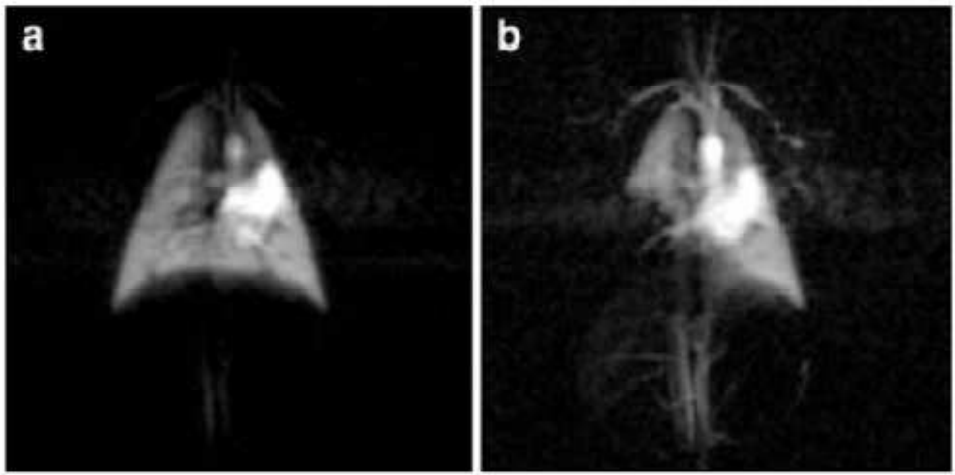

*Fig. 2.9*: Immagine $^{13}$C trueFISP dei polmoni di maiale ottenuta con 2-idrossietilacrilato PHIP, tramite iniezione in vena (a) e con catetere (b).

Accanto a tecniche già consolidate come MIGET (multiple inert-gas-elimination), SPECT e PET, si potrebbero potenzialmente sfruttare molecole iperpolarizzate $^{13}$C per ottenere mappe di perfusione ad alta risoluzione combinate a dati di ventilazione, ricavabili da oxygen-enhanced MRI o Xe-enhanced CT. Nell'analisi delle malattie polmonari è infatti molto importante la determinazione del rapporto ventilazione/perfusione (VA/Q), che è oggetto dei primi esami nei pazienti con sospetto embolismo polmonare. Il fatto che i nuclei iperpolarizzati sono spin in una situazione lontano dall'equilibrio rende inoltre possibile l'applicazione di una nuova tecnica di indagine, detta "bolus differentiation", basata sul fatto che la polarizzazione dei marker iperpolarizzati può essere cancellata applicando un'opportuna radiofrequenza di eccitazione. I principali vantaggi sono che la quantificazione è indipendente dal ritardo e dalla dispersione nelle arterie del marker e che il metodo può essere potenzialmente applicato a qualsiasi tessuto. In uno studio condotto su rene di coniglio da Golman et al. nel 2004 [3] è stato infatti dimostrato che con il ripetuto invio di rf depolarizzanti il mezzo di contrasto in una circoscritta fetta dell'immagine è stato possibile misurare il flusso di sangue in uscita dal tessuto registrando il segnale a tempi successivi la depolarizzazione (Fig. 2.10).

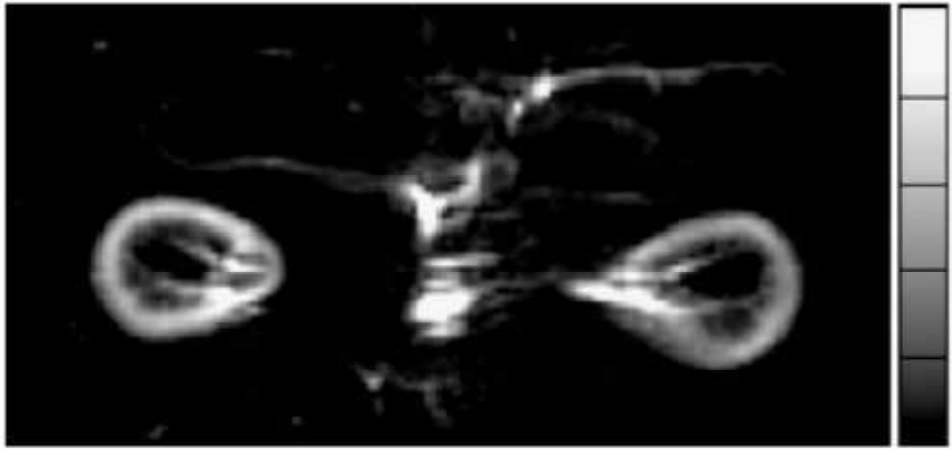

*Fig. 2.10*: Mappa di perfusione quantitativa della corteccia renale di un coniglio.

È stata ottenuta con la sequenza trueFISP che rapidamente distrugge la polarizzazione di una porzione, lasciando passare 1.5 s tra due immagini. Il 2-idrossietilacrilato è stato polarizzato tramite PHIP e iniettato via vena, 0.3 M. La perfusione stimata in questo modo è interessante perché mostra l'insensibilità verso la dispersione arteriosa e la possibilità di valutare l'influenza del rumore sul valore calcolato. Inoltre, si possono ricavare informazioni circa il tempo di transito e la dispersione del sangue nelle arterie che irrorano il tessuto. La somministrazione dei marker tramite cateteri arteriali è migliore, dato che il fattore di diluizione è minore. Le immagini ottenute in questo modo mostrano, infatti, un migliore SNR, al contrario di quanto accade per la somministrazione di classici agenti di contrasto per $^1$H MRI, perché alte concentrazioni di Gd tipicamente deviano la linearità segnale/concentrazione e riducono il SNR.

## 6.3 Catheter tracking

In generale, l'MRI presenta diversi vantaggi nelle esecuzioni di biopsie e la possibilità di registrare immagini multiplanari rende possibile la visualizzazione 3D del catetere. I metodi tradizionali possono però soffrire di alcune limitazioni legate a proprietà meccaniche, al riscaldamento del tessuto del catetere e ad una troppo piccola differenziazione di segnale tra il catetere e il tessuto circostante. L'assenza di rumore e l'alto segnale $^{13}$C iperpolarizzato rende tali molecole attraenti per procedure di interventi endovascolari MRI [3-5]. In questo caso però devono essere acquisite contemporaneamente alle immagini 13C anche immagini 1H, usando un sistema multinucleare, in modo da poter avere immagini anatomiche della regione di interesse e conoscere quindi l'esatta corrispondenza geometrica tra la posizione del catetere e l'intorno. Ad esempio, Golman et al. hanno riportato la visualizzazione di due cateteri attraverso l'aorta e l'arteria renale di un maiale (Fig. 2.11) [3]. Il 2-idrossietilacrilato iperpolarizzato via PHIP è stato iniettato, da un canale diverso dal catetere, nell'arteria e flussato fino al rene. Questo in teoria dovrebbe anche contribuire alla valutazione dello stato escretore del rene.

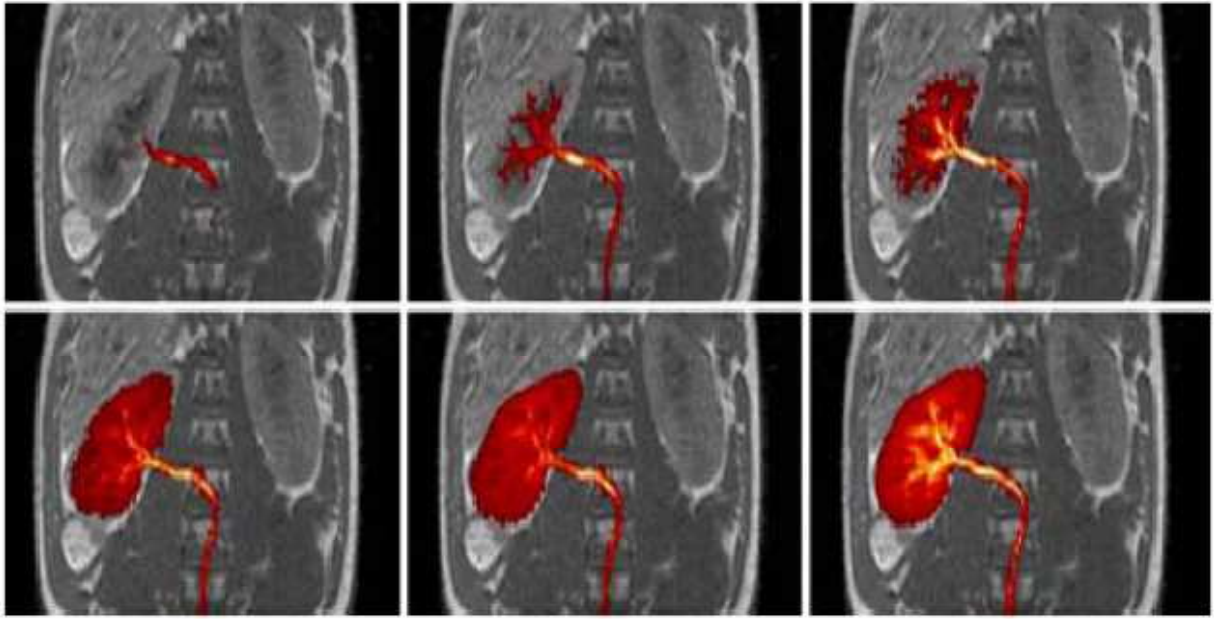

*Fig. 2.11*: Catheter tracking dell'arteria renale di maiale, ottenuto grazie all'iniezione di una soluzione di 2-idrossietilacrilato polarizzato tramite PHIP.

## 6.4 Imaging molecolare

L'imaging molecolare, cioè la capacità di visualizzare *in vivo* nell'uomo dei processi molecolari, si avvale di diverse tecniche della medicina nucleare, come PET, SPECT e MRI, la cui caratteristica comune è quella di usare specifici probes molecolari come sorgente di contrasto molecolare. È una disciplina moderna, che consente la visualizzazione di funzioni cellulari e metaboliche al livello biochimico più elementare. I targets includono i recettori delle superfici cellulari, gli enzimi di trasporto intercellulare, l'RNA messaggero, specifici tessuti, ecc. Il segnale per l'immagine può avere origine sia dalle molecole bersaglio che dai loro surrogati.

Alcune immagini *in vivo*, ottenute in seguito all'iniezione di agenti di contrasto $^{13}$C iperpolarizzati, come ad esempio il 2-idrossietilacrilato ottenuto tramite PHIP con polarizzazione del 30%, mostrano la distribuzione dell'agente di contrasto in vari organi di coniglio (Fig. 2.12) [5]. In particolare, è possibile seguire la distribuzione delle molecole $^{13}$C nel tempo, registrando un'immagine trueFISP per ogni coniglio a diversi tempi dall'iniezione. Le immagini mostrano che dopo 2 secondi la molecola è principalmente localizzata nel cuore e nei polmoni, dopo 4 secondi il segnale nei polmoni e nell'aorta diminuisce, mentre aumenta quello nei reni e nelle pareti dello stomaco e dopo 6 secondi si riesce a registrare anche un'immagine dell'intestino, a differenza delle immagini 1H che non sono in grado di mostrare né lo stomaco né l'intestino.

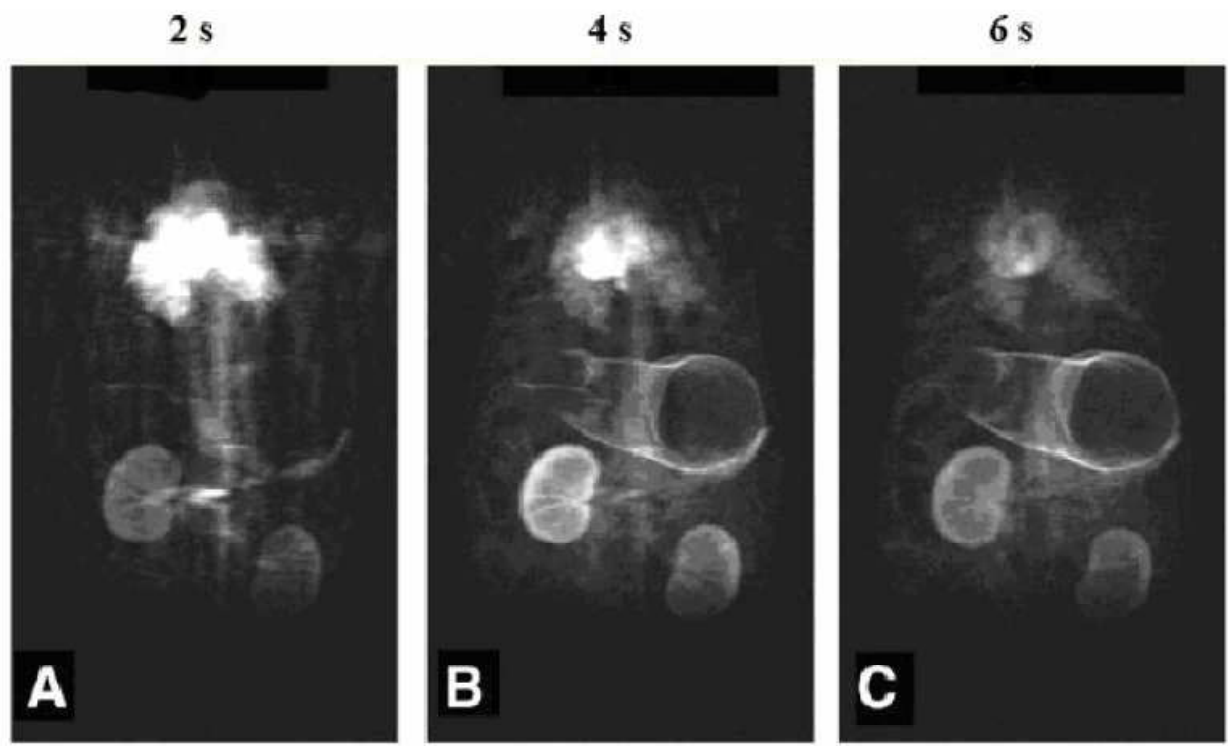

*Fig. 2.12*: Distribuzione dell'agente iperpolarizzato in vari organi di coniglio a diversi tempi dall'iniezione.

Inizialmente la distribuzione della molecola marcata $^{13}$C è molto simile a quella del convenzionale agente di contrasto paramagnetico, come è mostrato dall'alto segnale nel cuore, polmoni, rene e aorta. Ma già dopo 4 secondi dall'iniezione altre strutture, come le pareti dello stomaco, sono visibili solo nelle immagini $^{13}$C. Infatti, nei primi pochi secondi dopo l'iniezione la maggior parte dell'agente di contrasto $^{13}$C è ancora presente nel sangue e negli organi più vascolarizzati, come cuore, polmoni e reni, ma dato il suo basso peso molecolare, circa 120 g mol-1, esce rapidamente nello spazio extravascolare e dopo pochissimi secondi mostra un uptake significativo nei tessuti molli. Si pensa, ma non è ancora stato sperimentato, che sia possibile ottenere buone mappe di distribuzione iniettando e analizzando contemporaneamente più molecole $^{13}$C iperpolarizzate, riuscendo così ad ottenere maggiori informazioni circa la struttura delle membrane e la loro permeabilità.

Un'altra prospettiva molto interessante è l'iniezione di sostanze endogene iperpolarizzate $^{13}$C, velocemente metabolizzate dall'organismo. Mentre le tecniche PET e SPECT visualizzano solo la distribuzione dei nuclei radioattivi, senza discriminare i vari substrati marcati che contengono lo stesso nucleo radioattivo, la tecnica NMR è capace di distinguere il segnale derivante da molecole diverse, perché esso è strettamente correlato alla struttura molecolare e all'intorno chimico fisico. L'ampio range di chemical shift tipico del carbonio, più di 200 ppm, permetterebbe quindi di distinguere i segnali del reagente iniettato rispetto a quelli di suoi eventuali prodotti metabolici. Si inizia a parlare di Imaging Metabolico.